\documentclass[aps,pra,english,showkeys,amsmath,amssymb,twocolumn,showpacs]{revtex4}
\usepackage{subfigure}
\usepackage[T1]{fontenc}
\usepackage[latin1]{inputenc}
\usepackage{babel}%
\usepackage{graphicx}
\usepackage{color}
\usepackage{bm}
\usepackage{longtable}
\usepackage{amsmath}
\usepackage{amsfonts}
\usepackage{amssymb}

\begin{document}

\title{Quadrature Uncertainty and Information Entropy of Quantum Elliptical Vortex States}

\begin{abstract}
We study the quadrature uncertainty of the quantum elliptical vortex state using the associated Wigner function. Deviations from the minimum uncertainty states were observed due to the absence of the Gaussian nature. In our study of the entropy, we noticed that with increasing vorticity, entropy increases for both the modes. We further observed that, there exists an \emph{optimum value of ellipticity which gives rise to maximum entanglement} of the two modes of the quantum elliptical vortex states. A further increase in ellipticity reduces the entropy thereby resulting in a loss of information carrying capacity. We check the validity of the entropic inequality relations, namely the \emph{subaddivity} and the \emph{Araki-Lieb inequality}. The later was satisfied only for a very small range of the ellipticity of the vortex while the former seemed to be valid at all values.
\end{abstract}

\author{Anindya Banerji$^{1}$\footnote{anindya@hetc.ac.in}}
\author{Prasanta K. Panigrahi$^{2}$\footnote{pprasanta@iiserkol.ac.in}}
\author{Ravindra Pratap Singh$^{3}$\footnote{rpsingh@prl.res.in}}
\author{Saurav Chowdhury$^{1}$}
\author{Abir Bandyopadhyay$^{1}$\footnote{abir@hetc.ac.in (Author for correspondence)}}
\affiliation{$^{1}$Hooghly Engineering and Technology College, Hooghly 712103, India}
\affiliation{$^{2}$Indian Institute for Science Education and Research, Kolkata 741252, India}
\affiliation{$^{3}$Physical Research Laboratory, Ahmedabad 380009, India}

\date{\today}


\maketitle

\section{Introduction}

Many of the fundamental concepts of quantum mechanics have been adapted from classical mechanics, albeit with some modifications.
One such concept of particular importance is the phase space, which remains useful when passing to quantum mechanics.
Similar to the probability density distribution functions in classical mechanics, quasi-probability distributions were introduced
in quantum mechanics \cite{Intro}. Unlike the former, these can be negative which justifies the name, quasi-probability distribution functions. Among them, the Wigner function stands out because it is real, nonsingular and yields correct quantum
mechanical operator averages in terms of phase-space integrals and possesses positive definite marginal distributions \cite{Wigner}.
The Wigner distribution function has come to play an ever increasing role in the description of both coherent and and partially coherent beams
and their passage through first order systems \cite{Wigner2}. Once the Wigner function is known, other properties of the system can be calculated from it. \\
Another concept, which quantum mechanics derives from classical mechanics, is that of entropy. It is a natural extension of the classical concept
when dealing with quantum statistical mechanics. It is also a key concept in the field of quantum information theory. The entropy of quantum states,
described in terms of density operators, replacing the classical probability theory, is defined by von Neumann entropy \cite{Nielson}.
It gives a measure of how much uncertainty there is in the state of a physical system and also gives an idea about the information carrying capability
of a state which is the task of present study.\\
Optical vortices, possessing orbital angular momentum, have been studied classically, by using phase plate or computer generated hologram,
for quite some time \cite{Allen}. Agarwal \emph{et al} gave the concept of quantum optical vortex, which were circular
in shape \cite{GSA97,JBanerji,GSAJhaBoyd}. It was later generalized to quantum elliptical vortex (QEV) \cite{ABOpticsComm2011}.
In a recent work, it has been pointed out that photon substraction from one of the spontaneous parametric down converted beams (idler) produces
an elliptical vortex state \cite{NJP}. It should be highlighted that the photon subtraction from spontaneous parametric down converted light has
been realized experimentally as well \cite{Parigi,ArXiv}.\\
The paper is organized as follows. In section II we introduce the concept of QEV states in terms of
basis vector states. We write the corresponding state for an elliptical vortex. We briefly discuss
the Wigner distribution function for generalized quantum vortex. Using the same function, we
calculate the uncertainty products and discuss the results therein. We write down the corresponding
reduced density matrices for QEV states in section III and study the entropy of the constituent
modes. In section IV we verify the validity of the entropic inequalities, namely,
\emph{subadditivity} and \emph{Araki-Lieb} inequality for the QEV states. We conclude this article
after pointing out the significant results and directions for future work in section V.

\section{Uncertainties of the Two Modes of Quantum Elliptical Vortex (QEV) States}

The Gaussian wave packets occupy a central place in studies involving wave packets of quantum
system. In case of radiation fields these packets play an important role as these are in fact
minimum uncertainty states and describe both, the coherent states \cite{Glauber} as well as the
squeezed states \cite{Yuen}. 
%The quantum elliptical vortex states, as in the following equation,
%are of non-Gaussian structure
%\begin{eqnarray}
%\label{gauss}
%\psi_{QEV}(x,y)&\sim & (\eta_{x}x-i\eta_{y}y)^{m}\nonumber \\ \times &\exp & \left[-\frac{1}{2} \left\lbrace\left(\frac{x}{\sigma_{x}}\right)^{2}+\left(\frac{y}{\sigma_{y}}\right)^{2}\right\rbrace\right]
%\end{eqnarray}
%\noindent where \emph{m} is an integer, measuring vorticity, or the maximum OAM of the state.\\
The squeezed state $\vert\psi\rangle$ for the two mode radiation field is defined as the direct product of the
two squeezed mode states $\vert\psi\rangle_{a}$ and $\vert\psi\rangle_{b}$,
\begin{eqnarray}
\label{vortex}
\vert\psi\rangle&\equiv&\vert\psi\rangle_{a}\vert\psi\rangle_{b}\nonumber \\
&=&\exp\left[\zeta_{x}\lbrace a^{\dagger2}-a^{2}\rbrace\right]\vert0\rangle\nonumber\\
& &\times\exp\left[\zeta_{y}\lbrace b^{\dagger2}-b^{2}\rbrace\right]\vert0\rangle.
\end{eqnarray}
\noindent Here \emph{a}, \emph{b} are
the regular bosonic annihilation operators for the two modes and $\zeta_{i}$'s are the squeezing
parameters. Invoking the disentangling theorem \cite{Yuen}, the exponential for one mode in Eq.
(\ref{vortex}) can be expressed as
\begin{eqnarray}
\label{proexp}
&\exp &\left[\zeta_{x}\left(a^{\dagger2}-a^{2}\right)\right]=\exp\left[\frac{\xi_{x}}{2}a^{\dagger2}\right]\nonumber\\ &\times &
\exp \left\lbrace -\text{ln} [\text{cosh}(2\zeta_{x})]\left(a^{\dagger}a+\frac{1}{2}\right)\right\rbrace \nonumber \\
&\times & \exp\left[\frac{\xi_{x}}{2}a^{2}\right],\hspace{0.2cm} \xi_{x}=\text{tanh}(2\zeta_{x})
\end{eqnarray}
\noindent which is a product of exponentials. Using the fact that \emph{a} and \emph{b} acting on their respective vacuum states give zero,
we can write the squeezed state for mode \emph{a} as
\begin{equation}
\label{amode}
\vert\psi\rangle_{a}=\frac{1}{\sqrt{\text{cosh}(2\zeta_{x})}}\exp\left[\frac{\xi_{x}}{2}a^{\dagger2}\right]\vert0\rangle.
\end{equation}
\noindent Following a similar approach we can also write the corresponding expression for
$\vert\psi\rangle_{b}$. The vortex state can now be written as
\begin{equation}
\label{vortex1}
\vert\psi_{QEV}\rangle=A\left(\eta_{x}a^{\dagger}-i\eta_{y}b^{\dagger}\right)^{m}\vert\psi\rangle,
\end{equation}
\noindent where \emph{A} is the normalization constant, $\vert\psi\rangle$ represents the two
squeezed mode state as in Eq. (\ref{vortex}), \emph{m} stands for the vorticity and $\eta_{x}$,
$\eta_{y}$ control the ellipticity of the vortex. The spatial distribution associated with Eq. (\ref{vortex1}) has the following structure
\begin{eqnarray}
\label{gauss}
\psi_{QEV}(x,y)&\sim & (\eta_{x}x-i\eta_{y}y)^{m}\nonumber \\ \times &\exp & \left[-\frac{1}{2} \left\lbrace\left(\frac{x}{\sigma_{x}}\right)^{2}+\left(\frac{y}{\sigma_{y}}\right)^{2}\right\rbrace\right]
\end{eqnarray}
If we put $\eta_{x}=\eta_{y}=1$, $\zeta_{x}=\zeta_{y}=\zeta$ (real), it reduces to the circular vortex state (QOV) in a Gaussian beam. Using $\eta_{i}=1/\left(\sqrt{2}\sigma_{i}\right)$, the normalized spatial distribution of the QEV state \cite{ABOpticsComm2011} is obtained as
\begin{eqnarray}
\label{QEV}
\Psi_{QEV}(x,y) = \sqrt{\frac{2^{(m-2)}}{\sigma_{x}\sigma_{y}\Gamma(m+\frac{1}{2})\sqrt{\pi}}}\hspace{1cm}\nonumber\\
\times \left[\frac{x}{\sqrt{2}\sigma_{x}}\pm i\frac{y}{\sqrt{2}\sigma_{y}}\right]^{m}\nonumber \\
\times\exp \left[-\frac{1}{2}
\left\lbrace\left(\frac{x}{\sigma_{x}}\right)^{2}+\left(\frac{y}{\sigma_{y}}\right)^{2}\right\rbrace\right]
\end{eqnarray}
\noindent where $\sigma_{i}=\exp(2\zeta_{i})$. As is clear from Eq. (\ref{QEV}), the QEV states are non - Gaussian in structure.\\
We use $\sigma_{y}=\sqrt{5\sigma_{x}}$ or equivalently, $\zeta_y=\frac{\text{ln}5}{4}+\zeta_x$ arbitrarily and change the variables to a new set of scaled ones defined as $X_1=\frac{x}{\sigma_x}$, $Y_1=\frac{y}{\sigma_y}$, $X_2=\frac{\sigma_yx}{2\sigma_x}$, $Y_2=\frac{\sigma_xy}{2\sigma_y}$, $P_{X_{1}}=\frac{\sigma_{x}}{\sqrt{2}}p_{x}$, $P_{Y_{1}}=\frac{\sigma_{y}}{\sqrt{2}}p_{y}$,
$P_{X_{2}}=\frac{\sigma_{y}^{3}}{\sqrt{2}}p_{x}$ and $P_{Y_{2}}=\frac{\sigma_{x}^{3}}{\sqrt{2}}p_{y}$ \cite{ABOpticsComm2011}.
Following the treatments of \cite{RP2007}, the four dimensional Wigner function for the state $\Psi_{QEV}(x,y)$, is obtained
in a compact fashion as,
\begin{eqnarray}
\label{wigner2}
W\left(x,y,p_{x},p_{y}\right)\hspace{4cm} \nonumber\\
= K \exp\left[-\left(X_{1}^{2}+Y_{1}^{2}+P_{X_{1}}^{2}+P_{Y_{1}}^{2}
\right)\right]\nonumber \\
\times L_{m}^{-1/2}\left[\frac{\left(P_{X_{2}}+P_{Y_{2}}-X_{2}-Y_{2}\right)^2}{\sigma_{x}^{2}+\sigma_{y}^{2}}\right]
\end{eqnarray}
\noindent where $K=\frac{2^{m-4}m!}{\pi\sqrt{\pi}\Gamma(m+\frac{1}{2})}\left[-2\left(\sigma_{x}^{2}+\sigma_{y}^{2}\right)\right]^{m}$ and $L_{m}^{-1/2}$
is the associated Laguerre polynomial.\\
\begin{figure}[h]
\begin{center}
\includegraphics[scale=0.7]{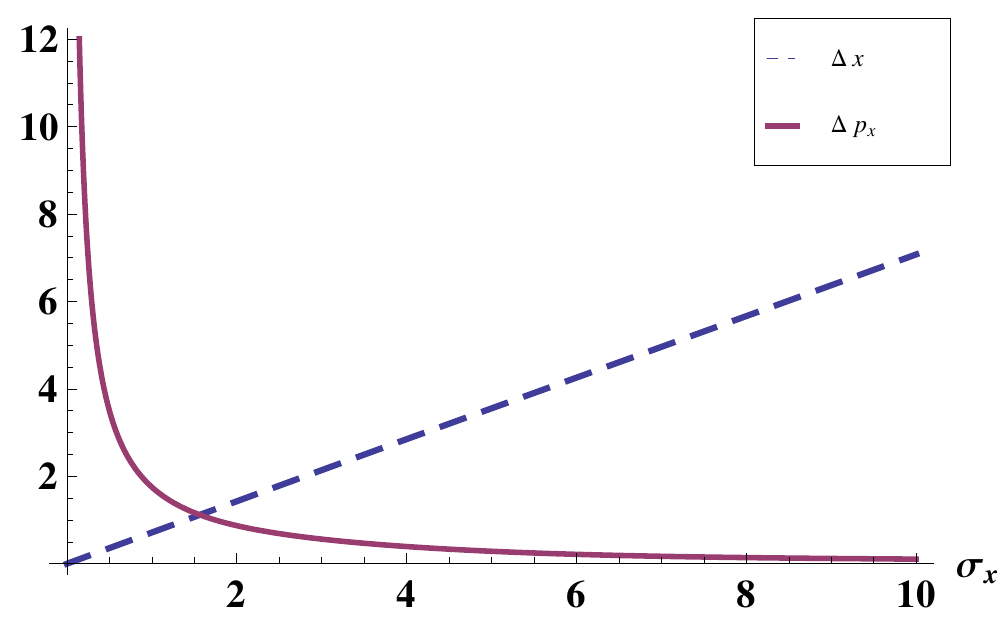}
\end{center}
\caption{(Color online) Variation of uncertainties in quadratures in one mode with $\sigma_{x}$ }
\label{fig:xpx}
\end{figure}
\begin{figure}[h!]
\begin{center}
\includegraphics[scale=0.7]{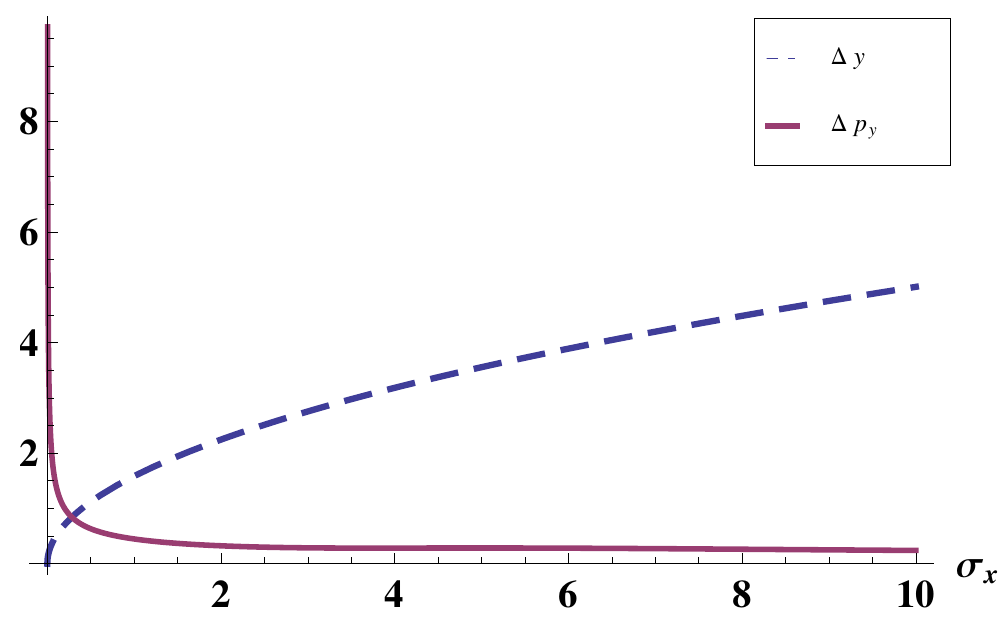}
\end{center}
\caption{(Color online) Variation of uncertainties in quadratures of the other mode with
$\sigma_{x}$ } \label{fig:ypy}
\end{figure}
\begin{figure}[h!]
\begin{center}
\includegraphics[scale=0.7]{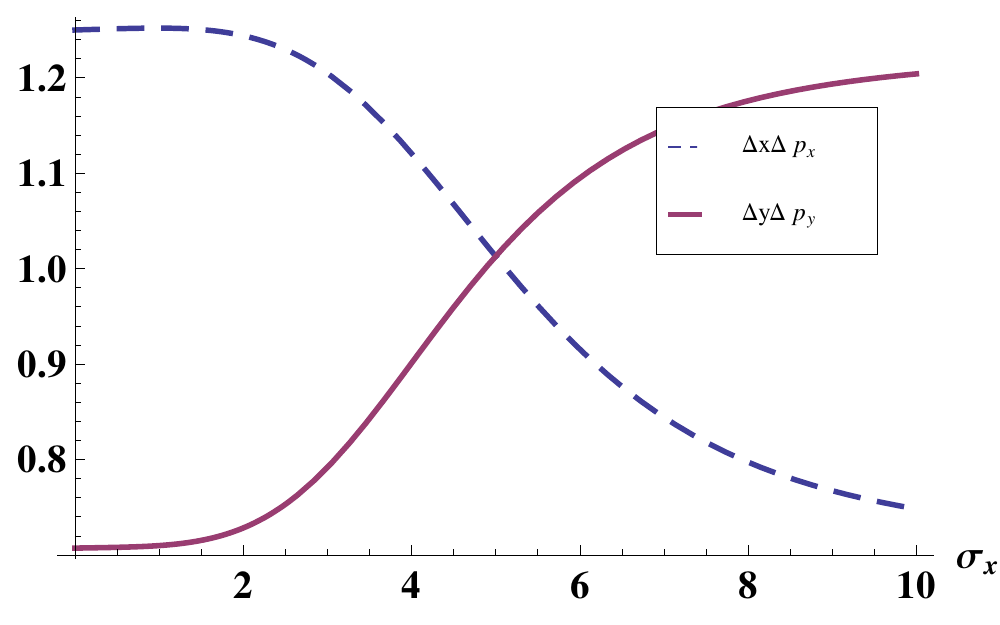}
\end{center}
\caption{(Color online) Uncertainty products of both the modes with $\sigma_{x}$}
\label{fig:uncertainty}
\end{figure}
Using Eq. (\ref{wigner2}) we determine the uncertainty in $x$, $y$, $p_{x}$ and $p_{y}$.
\begin{figure}[h!]
\begin{center}
\includegraphics[scale=0.7]{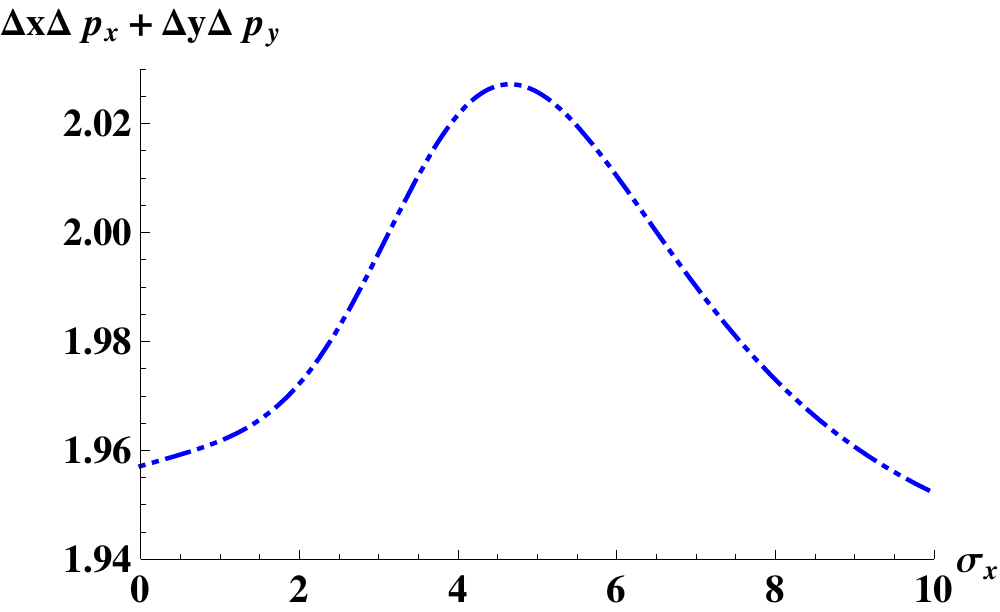}
\end{center}
\caption{(Color online) Sum of the uncertainty products with $\sigma_{x}$} \label{fig:sum}
\end{figure}
We study the uncertainties as a function of $\sigma_{x}$ for different vorticities \emph{m}. It is
seen from Fig. \ref{fig:xpx} that $\Delta x$ increases monotonically with increasing $\sigma_{x}$
whereas $\Delta p_{x}$ starts from infinity and falls to a much lower value with increasing
$\sigma_{x}$. This is expected as $\Delta x$ varies linearly with $\sigma_{x}$ and $\Delta p_{x}$
varies as $1/(\Delta x)$ and hence $1/(\sigma_{x})$. On the other hand $\Delta y$ increases
nonlinearly with $\sigma_{x}$ (Fig. \ref{fig:ypy}). This is mainly because of the parametrization
that we used for our calculations where $\Delta y$ varies as $\sqrt{5\sigma_{x}}$. $\Delta p_{y}$
on the other hand starts from infinity and falls as $1/\sqrt{\sigma_{x}}$, thus preserving the
Heisenberg uncertainty relation. It is further observed that at a particular value of $\sigma _{x}$
all the uncertainties are same. It is noticed from Fig. \ref{fig:uncertainty} that it is no longer
the minimum uncertainty state because the QEV state is not a ground state but an excited state unlike the two squeezed mode vacuum. Moreover, as the two modes are entangled, our results show that when the uncertainty product of one mode is decreased, the other one is increased. Hence the uncertainty product $\Delta x\Delta p_{x}$ decays to $1/\sqrt{2}$. $\Delta y\Delta p_{y}$ has an almost complementary nature to that of the former. It starts from around $1/\sqrt{2}$ and gradually increases to reach saturation at a much higher value. The maximum value attained by $\Delta x \Delta p_{x}$ is different than that attained by $\Delta y\Delta p_{y}$. The same holds true for the minimum values as well. However, it is observed that neither of the uncertainty products violate the uncertainty principle. In Fig. \ref{fig:sum} we study
the sum of the uncertainty products, $\Delta x \Delta p_{x} + \Delta y \Delta p_{y}$. 

\section{Reduced state for each mode and information entropy}

In this section we calculate the reduced density matrices of the two modes and calculate the
corresponding entropies. We start with determining the expression for the density matrix of the QEV
state $\vert\psi_{QEV}\rangle$. Using Eq. (\ref{amode}) and (\ref{vortex1}), one can write down the
vortex state in an expanded form in Fock state basis, as
\begin{eqnarray}
\label{vortex2}
\vert\psi_{QEV}\rangle &=& A\left(\eta_{x}a^{\dagger}-i\eta_{y} b^{\dagger}\right)^{m}\vert\psi\rangle_{a}\vert\psi\rangle_{b} \nonumber \\
&=& \frac{A}{\sqrt{\text{cosh}(2\zeta_{x})\text{cosh}(2\zeta_{y}}}\nonumber \\
&\times &\sum_{k=0}^{m} \frac{m!}{k!(m-k)!}\left(-i\eta_{y} \right)^{k}\eta_{x}^{m-k} \nonumber \\
&\times &\exp\left(\frac{\xi_{x}}{2}a^{\dagger2}\right)a^{\dagger ^{(m-k)}}\vert0\rangle_{a} \nonumber \\ &\times &\exp\left(\frac{\xi_{y}}{2}b^{\dagger2}\right)b^{\dagger ^{k}}\vert0\rangle_{b}.
\end{eqnarray}
\noindent The density matrix $\rho$ of the vortex state is
\begin{equation}
\label{densevor1} \rho = \vert\psi_{QEV}\rangle\langle\psi_{QEV}\vert.
\end{equation}
\noindent Using the properties of creation operators $a^{\dagger}$ and $b^{\dagger}$ and their
conjugates we can write down Eq. (\ref{densevor1}) in the following form
\begin{eqnarray}
\label{densevorfin}
\rho &=& \frac{A^{2}}{\text{cosh}(2\zeta_{x})\text{cosh}(2\zeta_{y})} \nonumber \\
&\times &\sum_{k=0}^{m}\frac{m!^{2}}{k!(m-k)!}\eta_{x}^{2(m-k)}\eta_{y}^{2k}\nonumber \\
&\times &\exp\left(\frac{\xi_{x}}{2}a^{\dagger2}\right)\vert m-k\rangle_{a}~_{a}\langle m-k\vert\exp\left(\frac{\xi_{x}}{2}a^{2}\right) \nonumber \\
&\times & \exp\left(\frac{\xi_{y}}{2}b^{\dagger2}\right)\vert k\rangle_{b}~_{b}\langle k\vert\exp\left(\frac{\xi_{y}}{2}b^{2}\right)
\end{eqnarray}
\noindent To calculate the entropy for each mode we need to obtain the corresponding reduced density matrices,
which is defined by $\rho_{a}\equiv \text{Tr}_{b} (\rho)$. In this context we should mention that the reduced state is a mixed state
even though the two mode state of Eq. (\ref{vortex}) is a pure state. Tracing out the \emph{b} mode, we can write down $\rho_{a}$ as
\begin{eqnarray}
\label{reduced}
\rho_{a} &=& \frac{A_{x}^{2}}{\text{cosh}(2\zeta_{x})\text{cosh}(2\zeta_{y})} \nonumber \\
&\times &\sum_{k=0}^{m}\frac{m!^{2}}{k!(m-k)!}\eta_{x}^{2(m-k)}\eta_{y}^{2k}\nonumber
\\ &\times &\exp\left(\frac{\xi_{x}}{2}a^{\dagger2}\right)\vert m-k\rangle_{a}~_{a}\langle m-k\vert\exp\left(\frac{\xi_{x}}{2}a^{2}\right)\nonumber \\
&\times & \left[_{b}\langle k\vert\exp\left(\frac{\xi_{y}}{2}b^{2}\right)\exp\left(\frac{\xi_{y}}{2}b^{\dagger2}\right)\vert k\rangle_{b}\right]
\end{eqnarray}
\noindent where $A_{x}$ is the normalization constant for mode \emph{a}. Solving the term in square bracket and rearranging we obtain
\begin{eqnarray}
\label{reducedfin}
\rho_{a} &=& \frac{A_{x}^{2}}{\text{cosh}(2\zeta_{x})\text{cosh}(2\zeta_{y})} \nonumber \\
&\times &\sum_{k=0}^{m}\frac{m!^{2}}{k!(m-k)!}\eta_{x}^{2(m-k)}\eta_{y}^{2k} \nonumber
\\ &\times &\exp\left(\frac{\xi_{x}}{2}a^{\dagger2}\right)\vert m-k\rangle_{a}~_{a}\langle m-k\vert\exp\left(\frac{\xi_{x}}{2}a^{2}\right)\nonumber \\
&\times & _{2}\text{F}_{1}\left[\frac{k+1}{2},\frac{k+2}{2},1,\xi_{y}^2\right]
\end{eqnarray}
\noindent where $_{2}\text{F}_{1}\left[\frac{k+1}{2},\frac{k+2}{2},1,\xi_{y}^2\right]$ is the Hypergeometric function \cite{Gradshteyn}. Using Eq. (\ref{reducedfin}) we calculate the diagonal elements of $\rho_{a}$. We make use of the fact that, $\text{Tr} \rho_{a}=\sum_{k=0}^{m}C_{k}^{a}$ where $C_{k}^{a}$ are the diagonal elements of $\rho_{a}$.\\
On rearranging the terms we finally arrive at the following form
\begin{eqnarray}
\label{coeff}
\text{Tr} \rho_{a} &=& \sum_{k=0}^{m}C_{k}^{a} \nonumber \\
&=& \sum_{k=0}^{m}\frac{A_{x}^{2}}{\text{cosh}(2\zeta_{x})\text{cosh}(2\zeta_{y})} \nonumber \\
&\times &\eta_{x}^{2(m-k)}\eta_{y}^{2k}\frac{m!^{2}}{(m-k)!}\nonumber \\ &\times & \sum_{n=0}^{\infty}\left(\frac{\xi_{x}}{2}\right)^{2n}\frac{1}{n!^{2}}\frac{(m-k)!}{(m-k-2n)!}\nonumber \\
\times \langle &m-&k-2n\vert m-k\rangle \langle m-k\vert m-k-2n\rangle \nonumber \\
&\times & _{2}\text{F}_{1}\left[\frac{k+1}{2},\frac{k+2}{2},1,\xi_{y}^2\right]
\end{eqnarray}
\noindent Invoking the orthogonality criterion we see that only the $n=0$ term will remain while
all other terms will cancel. So we can write down the simplified form of Eq. (\ref{coeff})
\begin{eqnarray}
\label{fincoeff}
\text{Tr} \rho_{a} &=& \sum_{k=0}^{m}C_{k}^{a} \nonumber \\
&=&\sum_{k=0}^{m}\frac{A_x^{2}}{\text{cosh}(2\zeta_{x})\text{cosh}(2\zeta_{y})}\nonumber \\
&\times &\eta_{x}^{2(m-k)}\eta_{y}^{2k}\frac{m!^{2}}{(m-k)!k!}\nonumber \\
&\times & _{2}\text{F}_{1}\left[\frac{k+1}{2},\frac{k+2}{2},1,\xi_{y}^{2}\right]
\end{eqnarray}
\noindent Using these coefficients we calculate the von Neumann entropy of mode \emph{a}, following the treatment of \cite{JBanerji} as follows
\begin{eqnarray}
\label{VNenta}
S_{a}&=&-\sum_{k=0}^{m} C_{k}^{a} \log_2 C_{k}^{a}\nonumber \\
& = & -\sum_{k=0}^{m} \frac{A_x^{2}}{\text{cosh}(2\zeta_{x})\text{cosh}(2\zeta_{y})}\eta_{x}^{2(m-k)}\eta_{y}^{2k}\frac{m!^{2}}{(m-k)!k!}\nonumber \\
&\times & _{2}\text{F}_{1}\left[\frac{k+1}{2},\frac{k+2}{2},1,\xi_{y}^{2}\right]\nonumber \\
&\times & \log_2 \frac{A_x^{2}}{\text{cosh}(2\zeta_{x})\text{cosh}(2\zeta_{y})}\eta_{x}^{2(m-k)}\eta_{y}^{2k}\frac{m!^{2}}{(m-k)!k!}\nonumber \\
&\times & _{2}\text{F}_{1}\left[\frac{k+1}{2},\frac{k+2}{2},1,\xi_{y}^{2}\right]
\end{eqnarray}
\noindent where $C_{k}^{a}$ stands for the coefficients in (\ref{fincoeff}). The logarithm is taken to base 2 as is the norm for information entropy. \\
Following a similar approach we can determine the von Neuman entropy for the \emph{b} mode. Repeating the procedures
of Eq. (\ref{reduced}) and Eq.  (\ref{reducedfin}) with the only difference that we trace out the \emph{a} mode instead
of \emph{b}, we can write the reduced density matrix $\rho_{b}\equiv\text{Tr}_{a}(\rho)$ as
\begin{eqnarray}
\label{reducedb}
\rho_{b} &=& \frac{A_{y}^{2}}{\text{cosh}(2\zeta_{x})\text{cosh}(2\zeta_{y})} \nonumber \\
&\times &\sum_{k=0}^{m}\frac{m!^{2}}{k!(m-k)!}\eta_{x}^{2(m-k)}\eta_{y}^{2k} \nonumber \\
&\times &\exp\left(\frac{\xi_{y}}{2}b^{\dagger2}\right)\vert k\rangle_{b}~_{b}\langle k\vert\exp\left(\frac{\xi_{y}}{2}b^{2}\right)\nonumber \\
&\times & _{2}\text{F}_{1}\left[\frac{m-k+1}{2},\frac{m-k+2}{2},1,\xi_{x}^2\right] \hspace{0.7cm}
\end{eqnarray}
\noindent where $A_{y}$ is the corresponding normalization constant for mode \emph{b}. Using Eq. (\ref{reducedb}) we determine the diagonal
elements of the reduced density matrix for the \emph{b} mode as follows
\begin{eqnarray}
\label{fincoeffb}
\text{Tr} \rho_{b} &=& \sum_{k=0}^{m}C_{k}^{b} \nonumber \\
&=&\sum_{k=0}^{m}\frac{A_y^{2}}{\text{cosh}(2\zeta_{x})\text{cosh}(2\zeta_{y})}\eta_{x}^{2(m-k)}\eta_{y}^{2k}\nonumber \\ &\times &
\frac{m!^{2}}{(m-k)!k!}\nonumber \\ &\times & _{2}\text{F}_{1}\left[\frac{m-k+1}{2},\frac{m-k+2}{2},1,\xi_{x}^{2}\right]
\end{eqnarray}
The corresponding entropy, $S_{b}$, can be calculated using Eq. (\ref{fincoeffb}) and the equivalent form of Eq. (\ref{VNenta}) by replacing \emph{a} with \emph{b}.\\
\begin{figure}[h!]
\begin{center}
\includegraphics[scale=0.7]{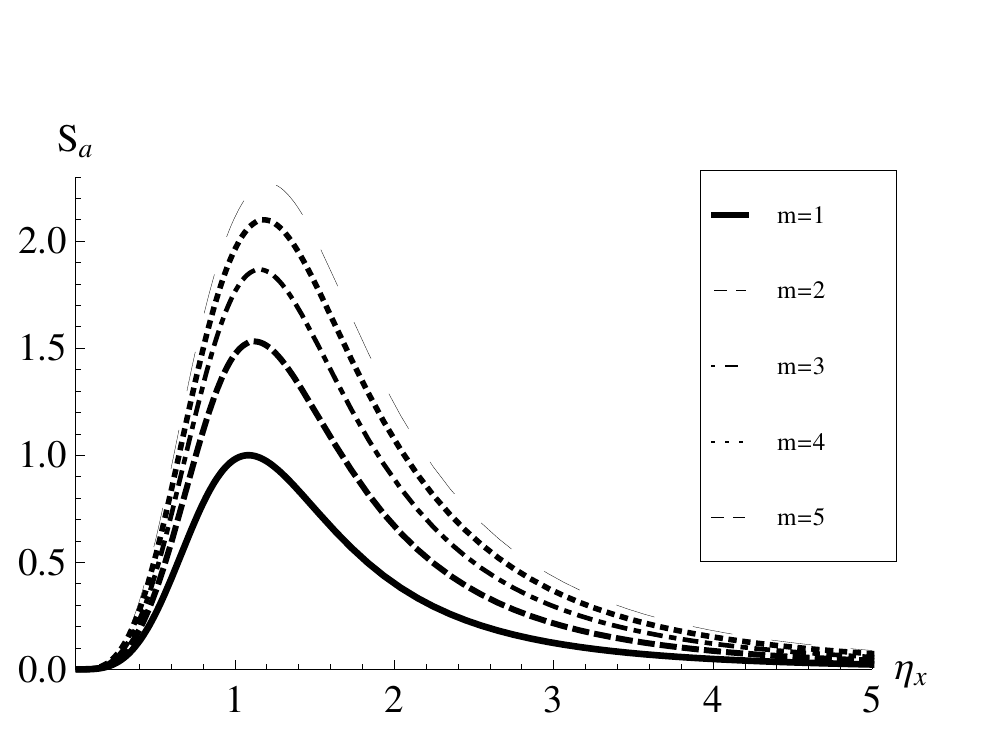}
\end{center}
\caption{(Color online) Variation of Entropy($S_{a}$) with $\eta_{x}$ } \label{fig:m1}
\end{figure}
We study the corresponding entropies with respect to $\eta_{x}$. We use $\sigma_{y}=3$,
$\sigma_{x}=5$ in our calculations for generating the graphs. We use the same parameterizations for
$ \zeta_{i}$ and  $\xi_{i}$ as in section II, where the subscript \emph{i} stands for \emph{x} and
\emph{y}. We choose $\eta_{y}=1/(\sqrt{2}\eta_{x})$ arbitrarily. The normalization constant is
evaluated for the two modes as
\begin{equation}
A_{i}^{2}=\frac{2^{2-k}\eta_{x}^{2(m-2k)}\sigma_{x}\sigma_{y}m!^{2}F_{i}}{(1+\sigma_{x}^2)(1+\sigma_{y})^{2}k!(m-k)!}
\end{equation}
\noindent where the subscript \emph{i} is used to denote the modes \emph{a} and \emph{b}. $F_{i}$ represents the corresponding Hypergeometric function as expressed in Eq. (\ref{reducedfin}) for mode \emph{a} and in Eq. (\ref{reducedb}) for mode \emph{b}.\\
\begin{figure}[h!]
\begin{center}
\includegraphics[scale=0.7]{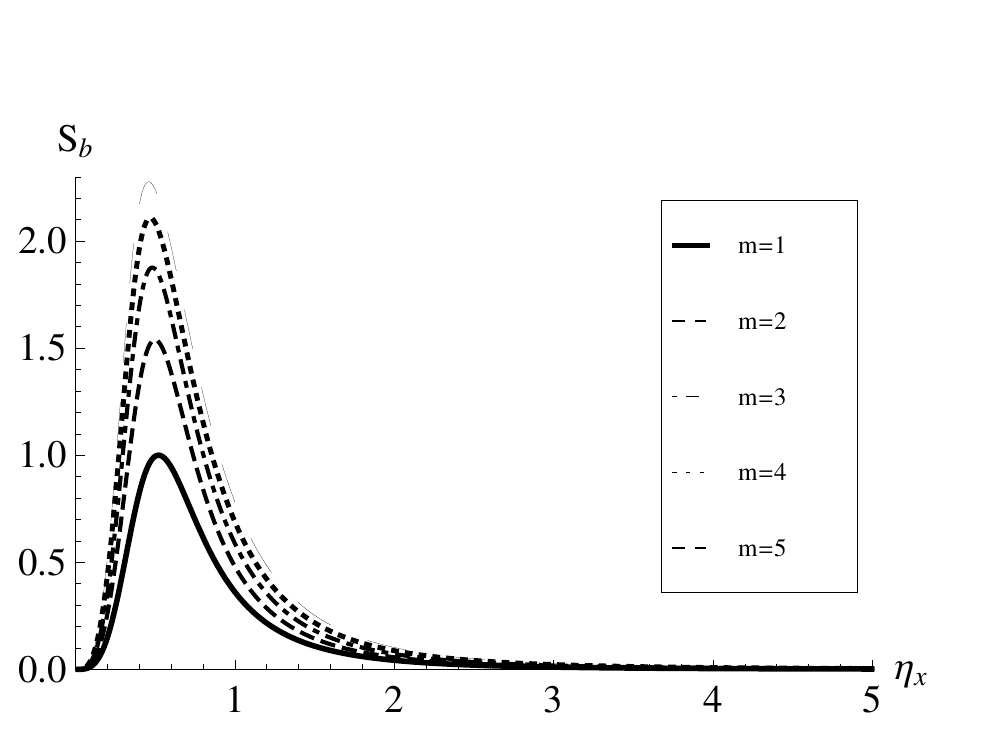}
\end{center}
\caption{(Color online) Variation of Entropy($S_{b}$) with $\eta_{x}$ } \label{fig:m2}
\end{figure}
We have studied the entropies of the two modes as functions of $\eta_{x}$ in Fig. \ref{fig:m1} and
Fig. \ref{fig:m2} for different \emph{m}, the vorticity. We observed that the entropy
increases with increasing vorticity. It is also observed that the peaks occurred at almost a fixed
value for each mode for all values of \emph{m} although it occurred for different values of
$\eta_{x}$ for the modes \emph{a} and \emph{b}. This signifies that there exists an optimum value
of $\eta_{x}$ for which maximum entanglement can be achieved for the QEV state. In other words,
\emph{an optimum level of ellipticity exists for which we can obtain the maximum entanglement}. We
use the word entanglement as information entropy is a measure of the degree of entanglement of the
constituent states which make up the system of states.\\
A further look at Fig. 5 and Fig. 6 makes us notice another very interesting fact. The maximum entropy occurs, not at the value of $\eta_{x}=1$, but at some other value. It is to be noted that $\eta_{x}=1$ corresponds to the circular vortex, which has been a topic of study for long. This observation leads us to conclude that \emph{elliptical vortex has more entropy than the circular vortex} which means more information transfer is possible by elliptical vortices, choosing correspondong ellipticity, than the circular vortex. This observation further emphasises the importance of the need to study elliptical vortices.\\
If the ellipticity is further increased,
entropy falls off exponentially. Thus we can conclude that, increasing the elliptic nature of the
vortex beyond a certain value would lead to disentanglement and hence loss of information carrying
capacity.

\section{Entropic inequalities and their validity for QEV states}

In the previous section we have calculated the reduced density matrices for the two modes using the partial trace operation and also calculated
the corresponding entropies. In this section we check the validity of the entropic uncertainty relations for the QEV state.\\
If two systems \emph{a} and \emph{a} have a joint quantum state $\rho_{ab}$, then the entropy of the combined states are expected to satisfy
the following inequalities \cite{Nielson}
\begin{eqnarray}
\label{Subadditivity}
S_{ab}&\le & S(a)+S(b)\\
\label{Araki-Lieb}
S_{ab}&\geq & |S(a)-S(b)|
\end{eqnarray}
\begin{figure}[h!]
\begin{center}
\includegraphics[scale=0.7]{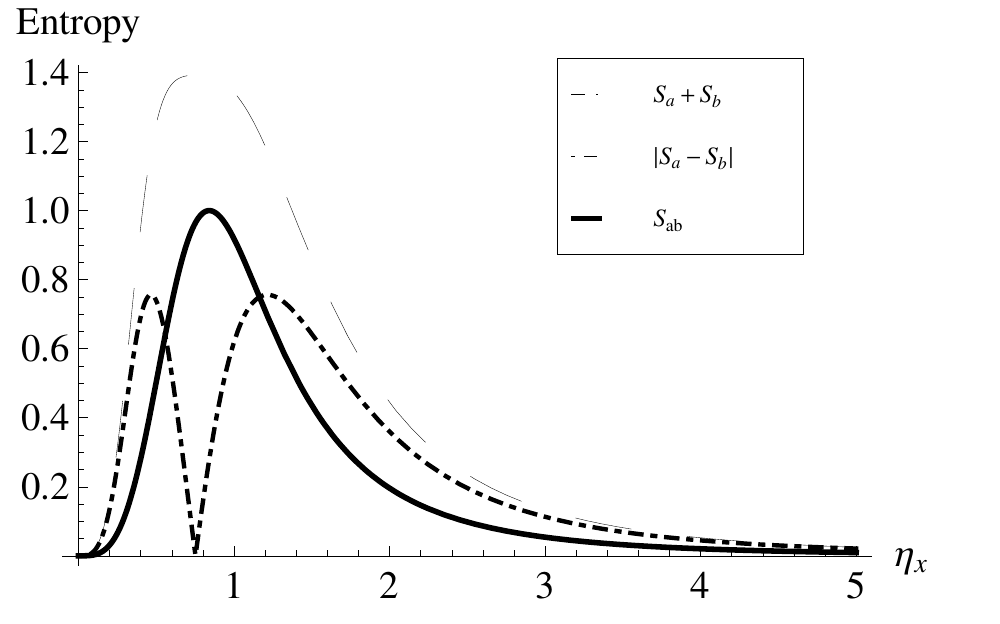}
\end{center}
\caption{Variation of Entropy with $\eta_{x}$ for $m=1$}
\label{fig:uncrtn1}
\end{figure}
We adapt these equations to suit our system where these are equally valid as the subsystems of the QEV state, $\rho_{a}$ and $\rho_{b}$ are
distinct quantum states though they are correlated. So we expect the \emph{Subadditivity}, Eq. (\ref{Subadditivity}) to hold with the inequality.
The second equation, i.e. Eq. (\ref{Araki-Lieb}) is the \emph{Araki-Lieb} inequality. It is generally satisfied for the von Neumann entropy. For a pure state, it signifies that entropy is cancelled only by an equal amount of entropy. But for mixed states, $S_{ab}>0$, so the entropies of the subsystems do not cancel each other completely. However, it is generally expected to be fulfilled for mixed states as well. This inequality also has some important implications for the index of correlation \cite{Barnett}. The index of correlation, $I_{c}$, is a measure of the information content of the correlation between the
components of an \emph{N} component system. For a two component system it can be expressed in a simplified form
\begin{equation}
\label{IndexCorrelation}
I_{c}=S_{a}+S_{b}-S_{ab}
\end{equation}
\noindent where $S_{ab}$ stands for the entropy of the combined system. If the two component system is in a pure state, it is shown that
maximum correlation occurs \cite{Lindblad,Phoenix}.
\begin{figure}[h!]
\begin{center}
\includegraphics[scale=0.7]{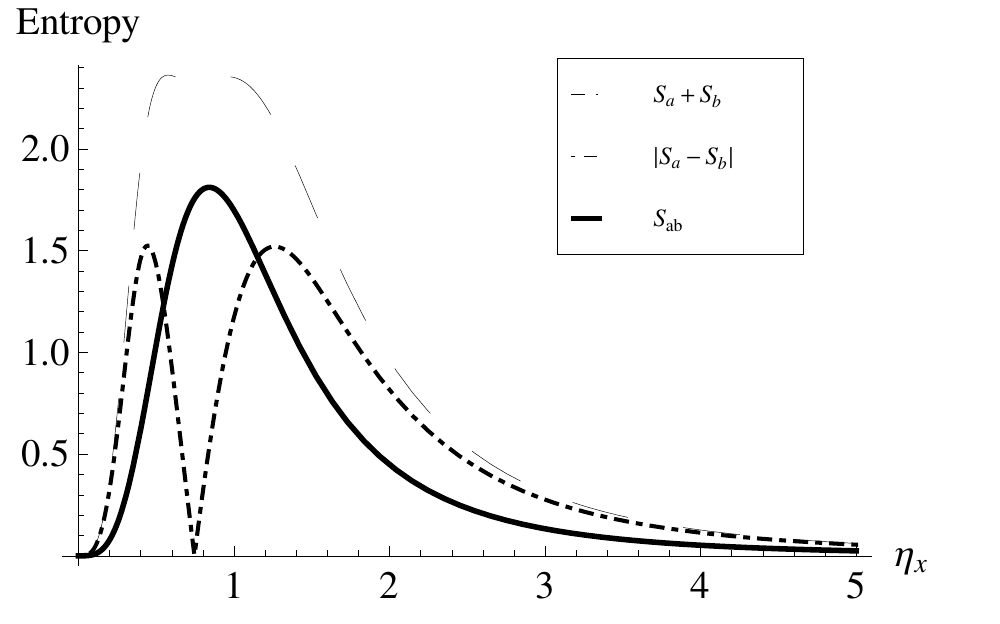}
\end{center}
\caption{Variation of Entropy with $\eta_{x}$ for $m=3$}
\label{fig:uncrtn3}
\end{figure}
The \emph{Araki-Lieb} inequality serves an important purpose here by limiting the maximum value that $I_{c}$ can take (Eq. (\ref{IndexCorrelation})),
thereby limiting the information content of the correlation of the components.\\
To verify these inequalities we need to determine the entropy of the combined system that is, both the modes taken together. In section III we have
written the density matrix of QEV state in Eq. (\ref{densevorfin}). We use it to calculate the von Neumann entropy of the state. Proceeding as we
did to determine the trace of the reduced density matrices, we can find the trace of the entire system as follows
\begin{eqnarray}
\label{CombinedSystemTrace}
\text{Tr} \rho &=& \sum_{k=0}^{m}C_{k}^{ab} \nonumber \\
&=& \frac{A^2}{\text{cosh}(2\zeta_{x})\text{cosh}(2\zeta_{y})}\nonumber \\
&\times & \sum_{k=0}^{m}\frac{m!^{2}}{(m-k)!k!}\eta_{x}^{2(m-k)}\eta_{y}^{2k}
\end{eqnarray}
Eq. (\ref{CombinedSystemTrace}) is used to determine entropy of the combined system, i.e., $S_{ab}$. We use the
same parametrization of the previous section to study them in Figs.
(\ref{fig:uncrtn1}-\ref{fig:uncrtn5}). It is observed that, $S_{a}+S_{b}$ starts from zero and
increases to reach saturation or a plateau region and falls off to zero. The joint entropy $S_{ab}$
on the other hand increases with increasing $\eta_{x}$ until it reaches a maximum. Then it starts
decreasing gradually to reach zero. $\vert S_{a}-S_{b}\vert$ exhibits a strikingly different nature
than the other two. It has a couple of peaks with the same maximum value while the other two attain
the maximum value only once. $\vert S_{a}-S_{b}\vert$ starts from zero and increases to attain the
maximum. After which it falls off to zero before rising again to the same maximum value for a
different value of $\eta_{x}$. Then it gradually falls off to zero like $S_{a}+S_{b}$ and $S_{ab}$.
The occurence of the zero at the middle can be explained very easily. It is clear from Fig.
\ref{fig:m1} and Fig. \ref{fig:m2} that both $S_{a}$ and $S_{b}$ attain the same value for that
particular
value of $\eta_{x}$ due to which $\vert S_{a}-S_{b}\vert$ goes to zero.\\
A further look at Figs. (\ref{fig:uncrtn1}-\ref{fig:uncrtn5}) enables us to infer that the
entropies satisfy the \emph{Subadditivity}, Eq. (\ref{Subadditivity}), for all values of $\eta_{x}$
but not with an equality as is expected for correlated systems. This signifies that Eq.
(\ref{Subadditivity}) holds for QEV states.
\begin{figure}[h!]
\begin{center}
\includegraphics[scale=0.7]{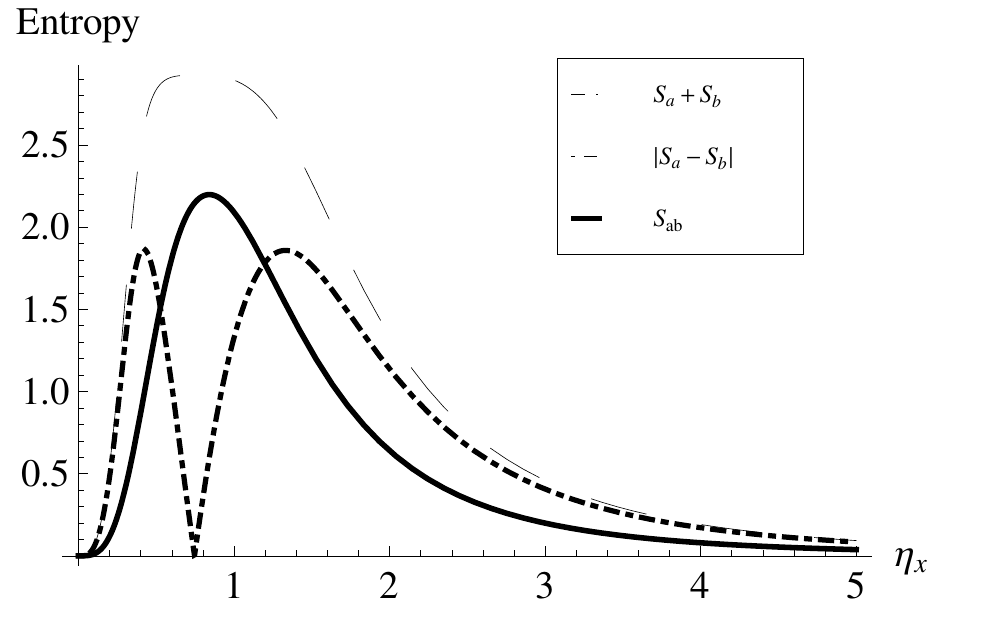}
\end{center}
\caption{Variation of Entropy with $\eta_{x}$ for $m=5$}
\label{fig:uncrtn5}
\end{figure}
This was also expected as the entropy for an entangled state should be less than that of the
summation of the entropies of the constituent systems. On the other hand the \emph{Araki-Lieb}
inequality, Eq. (\ref{Araki-Lieb}), was satisfied only in a very small range of values of
$\eta_{x}$. For all other values it is violated. We can argue that since the vortex is bounded by a
sharply peaked Gaussian distribution, it is in this region that Eq. (\ref{Araki-Lieb}) is violated
as verified by Keitel and Wodkiewicz \cite{Keitel}. More importantly, it is not fulfilled completely mainly because due to the correlation present between the two subsystems. As the subsystems \emph{A} and \emph{B} are entangled, the Araki - Lieb inequality is not fully fulfilled. Since $\eta_{x}$ controls the ellipticity of the vortex it is observed that the \emph{Araki-Lieb} inequality is valid only for a very short range of ellipticity. We, thus state, that the subsystems of the QEV state exhibit an optimum level of entanglement only in a limited range of the ellipticity of the QEV state where both the inequalities hold together. As the ellipticity increases $S_{ab}$ also increases and attains a maximum value where the inequality holds but as the ellipticity increases further the combined entropy falls off and the inequality is violated. This has some direct consequences which should be further investigated from a quantum information theoretic point of view to achieve maximum entanglement for this class of states.

\section{Conclusion}

In this paper we have calculated the uncertainty products using the Wigner function of the QEV
state. We noticed that the uncertainty product $\Delta x\Delta p_{x}$ attains a minimum value of 
$\frac{1}{\sqrt{2}}$. It has a maximum value of about 1.25. $\Delta y\Delta p_{y}$, on the other hand, has an initial value of about $1/\sqrt{2}$. It starts increasing gradually and saturates at nearly 1.2. It can be argued that the presence of the vortex modifies the characteristics of this state and hence it is no more the minimum uncertainty state which it would have been if the vortex was not present.\\
We have studied the von-Neumann entropy of the QEV states in terms of basis vector states. We found that the entropy was raised considerably with the increase in the vorticity of the states. It was noticed that the peaks for both the modes occur at different values of $\eta_{x}$ where $\eta_{x}$ is a measure of the ellipticity of the vortex. But the peak value of the entropies for the two modes remained the same. It was further observed that there exists an \emph{optimum value of ellipticity which gives rise to maximum entanglement} of the two modes of the QEV states. A further increase in ellipticity reduces the entropy thereby resulting in a loss of information carrying capacity.\\
We checked and verified the entropic inequalities. We observed that the strong subadditivity was satisfied for all conditions. This is expected for any entangled state, as the combined entropy for such state would be always less than that of the constituent systems taken together. We noticed that \emph{Araki-Lieb} inequality, an indicator of degree of entanglement, was \emph{violated} in all the regions, except in a very narrow region of ellipticity values.\\
Our results serve as a pointer for further investigation and studies of quantum elliptical vortices as means of information transport. The study of decoherence in this system needs to be pursued, for finding out the robustness of this correlated system. Investigation of quantum discord in this case will also be exciting, from both fundamental, as well as application point of view.

\section*{Acknowledgement}

This work is partially supported by DST through SERB grant no.: SR/S2/LOP-0002/2011.

\end{document}